\documentclass[12pt]{spieman}  
\usepackage{amsmath,amsfonts,amssymb}
\usepackage{graphicx}
\usepackage{setspace}
\usepackage{multirow}
\usepackage{algorithm2e}
\usepackage{tocloft}
\usepackage{lineno}

\title{SPIE journal papers: sample manuscript showing style and formatting specifications}

\cftpagenumbersoff{figure}
\cftpagenumbersoff{table}

\title{Self-optimizing adaptive optics control with Reinforcement Learning for high-contrast imaging}

\author[a,*]{Rico Landman}
\author[a,b,c]{Sebastiaan Y. Haffert}
\author[a]{Vikram M. Radhakrishnan}
\author[a]{Christoph U. Keller}

\affil[a]{Leiden Observatory, Leiden University, PO Box 9513, 2300 RA Leiden, The Netherlands}
\affil[b]{Steward Observatory, Unversity of Arizona, 933 North Cherry Avenue, Tucson, Arizona}
\affil[c]{NASA Hubble Fellow}

\begin{document}

\maketitle

\begin{abstract}
Current and future high-contrast imaging instruments require extreme adaptive optics (XAO) systems to reach contrasts necessary to directly image exoplanets. Telescope vibrations and the temporal error induced by the latency of the control loop limit the performance of these systems. One way to reduce these effects is to use predictive control. We describe how model-free Reinforcement Learning can be used to optimize a Recurrent Neural Network controller for closed-loop predictive control. First, we verify our proposed approach for tip-tilt control in simulations and a lab setup. The results show that this algorithm can effectively learn to mitigate vibrations and reduce the residuals for power-law input turbulence as compared to an optimal gain integrator. We also show that the controller can learn to minimize random vibrations without requiring online updating of the control law. Next, we show in simulations that our algorithm can also be applied to the control of a high-order deformable mirror. We demonstrate that our controller can provide two orders of magnitude improvement in contrast at small separations under stationary turbulence. Furthermore, we show more than an order of magnitude improvement in contrast for different wind velocities and directions without requiring online updating of the control law.
\end{abstract}


{\noindent \footnotesize\textbf{*}Rico Landman,  \linkable{rlandman@strw.leidenuniv.nl} }

\section{Introduction}
    One of the main limitations of current and future ground-based high contrast imaging (HCI) instruments is the wavefront error induced by the time lag between sensing and correcting the wavefront \cite{Guyon2018_XAO}. This time lag results in a halo of speckles along the wind direction, which limits the contrast at small separations \cite{Cantalloube2020_WDH_sphere}. Furthermore, the finite control bandwidth results in low-order residuals after the correction, such as tip and tilt. One of the main causes of these residuals are vibrations close to the maximum control frequency. All ground-based high-contrast imaging instruments suffer from vibrations, including SCExAO \cite{Lozi2018_vibration_scexao}, GPI \cite{Hartung2014_vibration_GPI} and SPHERE \cite{Sauvage2010_saxo}. These vibrations decrease the resolution of long-exposure images. Furthermore, some coronagraphs are highly sensitive to residual tip-tilt\cite{Lloyd2005_Lyot_tiptilt, Mawet2010_vector_vortex}, which will further reduce the contrast.
    
    Both of these effects can be reduced with better control algorithms. Early work on improved controllers involved optimizing the modal gains of a modal integrator based on the open-loop temporal Power Spectral Density (PSD)\cite{Gendron1995_OMGI_I,Gendron1995_OMGI_II}. This optimal gain is a trade-off between turbulence rejection and the amplification of noise and disturbances outside the control bandwidth. However, to reduce the effect of the servo-lag we have to predict the disturbances before they are measured. A popular approach to optimal control is Linear Quadratic Gaussian (LQG) control\cite{Paschall1993_LQG}, which is based on a Kalman filter. Variants of LQG-based controllers have been demonstrated to reduce tip-tilt vibrations in simulations \cite{Correia2012_vibrations}, lab verifications \cite{Petit2008_LQG_lab_vibration} and on-sky\cite{Sivo2014_LQG_canary, Petit2014_SPHERE_onsky, Hartung2014_vibration_GPI}. A disadvantage of LQG-like controllers is their reliance on a linear state-space model of the turbulence and deformable mirror (DM) dynamics\cite{Correia2010_DM_dynamics}; this requires accurate calibration of the model parameters. Hence, these controllers can also not correct for correlations in the turbulence that are not included in the model. A combination of an LQG controller for low-order vibrations and an optimal modal gain integrator for high-order modes is currently being used in the adaptive optics (AO) systems of both SPHERE \cite{Petit2014_SPHERE_onsky} and GPI \cite{Poyneer2016_GPI}.
    
    Another group of data-driven algorithms construct a linear filter that minimizes the residual phase variance from (pseudo) open-loop data. This idea was originally proposed and demonstrated for modal control by Dessenne et al. \cite{Dessenne1997}. More recently, a spatio-temporal filter based on Empirical Orthogonal Functions (EOFs) was demonstrated to significantly increase the contrast in a simulated HCI system\cite{Guyon2017_eof}. The performance of such linear filters was tested on open-loop AO telemetry for the Keck II AO system \cite{Jensen-Clem2019_EOF} and SPHERE\cite{VanKooten2020_PC} and showed a decrease in residual wavefront error. An issue with these data-driven predictive control algorithms is that they rely on the prediction of (pseudo) open-loop wavefront data, while XAO systems operate in closed-loop. To use these controllers, one has to reconstruct the pseudo open-loop residuals. Not only does this require knowledge of the servo-lag of the system and DM dynamics, it is also not trivial in the case where the response of the wavefront sensor is nonlinear, such as for the Pyramid Wavefront sensor \cite{Deo2019_optical_gain}. To circumvent the issue of nonlinearity, neural networks can improve the performance compared to linear models when using an unmodulated Pyramid Wavefront Sensor\cite{Wong2021_PC_NN}. However, predicting the open-loop residuals alone does not provide the optimal closed-loop control law. For example, one still has to find the best gain for the control loop. Furthermore, as the open-loop wavefront may be orders of magnitude larger than the closed-loop errors, a relative error in the predicted open-loop wavefront may be detrimental to the closed-loop contrast as compared with a relative error in the closed-loop residuals. Obtaining a closed-loop control law using supervised learning approaches is hampered by the distribution of inputs depending on the control law itself. It was recently shown that this can be mitigated by using an adversarial prior in the training of a neural-network controller\cite{Swanson2021_pc}. Instead of predicting in open-loop while controlling in closed-loop, we can also do data-driven predictive control directly in closed-loop. This can for example be done using data-driven subspace predictive control, which was recently shown to provide orders of magnitude improvement in contrast in simulations and in the lab\cite{Haffert2021_DDPSC}. This does not require the explicit prediction of the wavefront after a known servo-lag but directly optimizes the closed-loop commands. The proposed approach here is similar, but generalizes to nonlinear control laws and arbitrary optimization objectives.
    
   Another issue for most of these approaches is the non-stationary nature of turbulence \cite{VanKooten2019_nonstationary}. All methods mentioned above require online updating of the control parameters based on the current turbulence parameters. This updating of the control law is computationally very expensive and often involves inverting large matrices\cite{Gray2014_ensemble_LQG_nonstationary, Guyon2017_eof}. Recently, it was shown that a single Artificial Neural Network (ANN) can learn to predict the open-loop wavefront for varying wind speeds and directions for a simulated AO system\cite{Liu2020_LSTM}. This eliminates the need for updating the control law to keep up with changing turbulence properties. This would allow us to decouple the training of the controller from online control, simplifying the implementation and potentially gaining in stability. However, using this approach still leaves the problems associated with predicting the open-loop turbulence.
    
   Here, we use Reinforcement Learning (RL), a Machine Learning approach, for the data-driven optimization of the closed-loop adaptive optics control law. Indeed, Machine Learning has recently received significant attention in the AO community for wavefront reconstruction \cite{Landman2020, Norris2020_photonic_wfs}, prediction \cite{Liu2020_LSTM, Swanson2018_CNN_prediction} and control \cite{Wong2021_PC_NN, Swanson2021_pc}; Reinforcement Learning was already suggested as a method to directly optimize the contrast in a high-contrast imaging system \cite{Radhakrishnan2018} and for focal-plane wavefront control \cite{Sun2018_EM_FPWC}. Very recently, model-based RL was proposed for adaptive optics control \cite{Nousiainen_2021OExpr_RL}, and was shown to reduce the temporal error and adjust to mis-registrations between the deformable mirror and wavefront sensor. Our proposed approach here has similar advantages as their approach. The main difference is the use of a parameterized control law here over a planning algorithm. This avoids the need for solving the planning optimization for every control command, which is generally too computationally expensive to do at kHz frequencies, at which typical XAO systems are run. However, this generally comes at the cost of longer training times.
   
   Our RL-based approach has some distinct advantages:
    
    \begin{itemize}
        \item \textbf{Data-driven}:  The algorithm is completely self-optimizing and data-driven; it does not require an explicit model of the disturbances or system dynamics, or knowledge of system parameters. Instead of explicitly predicting the wavefront after an assumed servo-lag, it directly optimizes the closed-loop commands.
        
        \item \textbf{Flexible}: The algorithm allows for an arbitrary optimization objective. The user is free to choose the metric to optimize based on the science goals of the instrument. Furthermore, it can handle multiple inputs including systems with multiple wavefront sensors.
        
        \item \textbf{Nonlinear}: The proposed method can optimize any parameterized, differentiable control law. For example, one can optimize the parameters of a nonlinear Artificial Neural Network (ANN) controller. This allows the controller to model the nonlinear dynamics of the system, such as for systems with a Pyramid Wavefront Sensor. Because of the larger representation power of these Neural Networks, they can also learn to perform under varying atmospheric conditions.
        
        \end{itemize}
    This paper is structured as follows: Section \ref{sec:Methods} provides an introduction to Reinforcement Learning and describes the relevant algorithm. Section \ref{sec:tip_tilt_simulations} simulates the algorithm's performance in tip-tilt control for vibrations and a power-law input, while section \ref{sec:tip_tilt_lab} shows the results from a lab experiment. Section \ref{sec:full_wavefront} shows the results of using the algorithm to control a high-order deformable mirror. Finally, section \ref{sec:conclusions} draws conclusions and outlines future work.

\section{Reinforcement Learning Control}\label{sec:Methods}
    We begin the description of our algorithm with an introduction to the Reinforcement Learning framework, within which we formulate the AO control problem. This is followed by a discussion of the Deterministic Policy Gradient algorithm and how it can be applied to closed-loop AO control.
    
    \begin{table}[htbp]
    \caption{Overview and explanation of the various terms used in the Reinforcement Learning (RL) framework.}
    \label{tab:terms}
    \center
    \begin{tabular}{l|l|l}
    \hline
    \textbf{Symbol}  &    \textbf{RL term}           & \textbf{AO term/Explanation}             \\ \hline 
    $s$ & State   & Input to the controller.         \\ 
    $o$  & Observation & Measurement of the residual wavefront    \\ 
    $a$ &   Action   &  Incremental DM commands               \\
    $r$ & Reward      & Measure of the instantaneous performance; we are free to choose this \\
    $R$ & Return      & Discounted sum of future rewards; the optimization objective    \\
    $\pi_\theta$ & Actor/policy     & The control law                    \\ 
    $Q_\omega$ & Critic  &  Estimate of the return/the cost function \\ 
    \end{tabular}
    \end{table}
    
    \subsection{AO control as a Markov Decision Process}
    Reinforcement Learning\cite{sutton_barto_RL} (RL) algorithms deal with Markov Decision Processes (MDPs), discrete-time stochastic control processes. In the RL framework, an agent operates within an environment, which is formally defined by a tuple $(\mathcal{S}, \mathcal{A}, \mathcal{R}, \mathcal{P})$ and time index $t$. Here, $s_t \in \mathcal{S}$ is the state of the environment, $a_t \in \mathcal{A}$ the action taken by the agent, $\mathcal{R}$ the reward function and $\mathcal{P}$ the state transition probabilities defined by:
    \begin{equation} \label{eq:markov}
       \mathcal{P}(s_t,a_t,s_{t+1}) = P(s_{t+1}| s_t, a_t),
    \end{equation}
    where $P(s_{t+1}| s_t, a_t)$ gives the probability of transitioning from state $s_t$ to state $s_{t+1}$ as a result of action $a_t$. The state needs to be defined in such a way that the transition probabilities are fully described by Eq. \ref{eq:markov} and do not depend on previous states (e.g. $s_{t-1}$). This is also known as the Markov property.
    
    In the case of AO control, the agent is the DM controller. The environment consists of everything but the controller, including the evolution of the turbulence and the DM dynamics. At each time step the the controller receives a state $s_t$. For example, in the case of an integral controller the state consists of only the most recent observation of the residual wavefront $o_t$. The representation of the state for our Reinforcement Learning controller will be extensively discussed in section \ref{sec:state_representation}. Based on this state, the controller takes an action $a_t$, following some control law $\pi$: $a_t= \pi(s_t)$. Since the controller operates in closed loop, this action consists of the incremental voltages added to the DM, equivalent to an integral controller. The DM commands and the temporal evolution of the environment result in the transition to a new state $s_{t+1}$, according to the transition probabilities $\mathcal{P}$ of the stochastic process of the turbulence, and a reward $r_t = \mathcal{R}(s_t, a_t, s_{t+1})$. This reward is a measure of the performance of the controller; the reward function $\mathcal{R}$ can be arbitrarily chosen as long as it can be calculated at every time step. The goal is to determine the control law $\pi$ that gives the highest cumulative future reward. To ensure that the cumulative future reward remains finite and to prioritize immediate rewards, future rewards are discounted at a rate $\gamma<1$, the discount factor. The discounted future return $R_t$ is then defined as:
    \begin{equation}
        R_t = \sum_{t=0} \gamma^t r_t = r_t+\gamma r_{t+1}+ \gamma^2 r_{t+2} + \dots\;.
    \end{equation}
    The optimization objective can then be formally defined as:
    \begin{equation}
        J(\pi) = \mathbb{E}_{s\sim \rho^\pi, a\sim \pi}[R].
    \end{equation}
    Here, $\mathbb{E}$ denotes the expectation value and $\rho^\pi$ the state visitation distribution when following the control law $\pi$. This state visitation distribution describes how often we end up in state $s$ if we follow the control law $\pi$. This expectation over the state visitation distribution is taken because we want to optimize over the states that we observe. It also accounts for the dependency of the inputs on the controller. In the rest of this work, when we denote an expectation value, it is always over $s \sim \rho^\pi$ and $a \sim \pi$, and we will leave this out for readability. There are a variety of algorithms to find an optimal control law in the above framework\cite{Arulkumaran2017_RL_review}. One of the most frequently used algorithms for continuous control problems is the (Deep) Deterministic Policy Gradient algorithm \cite{Silver_dpg, Lillicrap2015_ddpg}, which we will use here. Our choice for the DDPG algorithm was motivated by the need for a directly parameterized and deterministic control law, which allows for the fast computation of the DM commands at kHz frequencies without needing to solve a planning optimization problem every millisecond. Furthermore, while not explored in this work, we preferred an algorithm which allows for an arbitrary optimization objective and could therefore also be used for the direct optimization of a focal plane quantity, such as the Strehl ratio or the post-coronagraphic contrast. Finally, since it is generally easy to collect a lot of data, as we generate 1000 datapoints per second at 1 kHz, the lower sample efficiency of model-free\footnote{Model-free in the RL context does not imply that we do not have an empirical model for the controller. Instead, it means that the algorithm does not explicitly model the transition dynamics and plans through this model to find the optimal commands/policy.} algorithms was considered less of an issue. 
    
    \subsection{(Deep) Deterministic Policy Gradient}
    The (Deep) Deterministic Policy Gradient (DPG)\cite{Silver_dpg} algorithm uses two parameterized models:
    \begin{itemize}
        \item An \textbf{Actor} $\pi_\theta(s)$ that models the control law with parameters $\theta$. This model maps the state to the DM commands : $a_t =\pi_\theta(s_t)$.

        \item A \textbf{Critic} $Q_\omega(s,a)$ that models the expected return for a given state and DM command with parameters $\omega$: $Q_\omega(s_t,a_t) = \mathbb{E}[R_t|s_t,a_t]$. This is an estimation of the cost function that has to be optimized.
    \end{itemize}
    Both the actor and critic may be any kind of parameterized, differentiable model. These may for example be function approximators, such as Artificial Neural Networks (ANN) \cite{Goodfellow_deeplearning}. The DPG algorithm with ANNs as function approximators is referred to as the Deep Deterministic Policy Gradient (DDPG) algorithm \cite{Lillicrap2015_ddpg}. In this case, $\theta$ and $\omega$ are the trainable parameters of the ANN for the actor and critic, respectively. The algorithm aims to find the parameters $\theta$ of the controller that maximizes the optimization objective $J(\pi_\theta)$. Using the actor and the critic we can calculate the gradients of the optimization objective with respect to the control parameters using the chain rule:
    \begin{equation}
    \begin{split}
         \nabla_\theta J(\pi_\theta) =& \mathbb{E}[\nabla_\theta Q_\omega(s,a)]\\
         =& \mathbb{E}[\nabla_a Q_\omega(s,a) \nabla_\theta\pi_\theta(s)]
    \end{split}
    \end{equation}
    These gradients can then be used by a gradient-based optimizer to update the parameters $\theta$ of the control law in the direction in which the expected return increases. The quality of the controller thus inherently depends on the ability of the critic $Q_\omega(s,a)$ to successfully model the expected return. The critic can be taught using Temporal Difference (TD) learning, which uses the Bellman equation\cite{bellman1954} to bootstrap the expected future reward:
    \begin{equation}\label{eq:bellman}
        \begin{split}
    Q_\omega(s_t,a_t) = & \mathbb{E}[R_t| s_t, a_t]\\
    = &\mathbb{E}[r_t + \gamma r_{t+1} + \gamma^2 r_{t+2}+\dots] \\
              = &\mathbb{E}[r_t+\gamma (r_{t+1}+\gamma r_{t+2}+\dots)] \\
              = &\mathbb{E}[r_t+\gamma Q_\omega(s_{t+1},\pi(s_{t+1}))] \equiv \mathbb{E}[y_t],
        \end{split}
    \end{equation}
    where $y_t$ are referred to as the target values and are calculated using:
    \begin{equation}
        y_t = r_t+\gamma Q_\omega(s_{t+1},\pi(s_{t+1})).
    \end{equation}
    Training the critic can then be framed into a supervised learning problem for each iteration, where we try to find the parameters $\omega$ that minimize the mean-squared error between the output of the critic and the target values $y_t$. The critic loss $L$ is thus given by:
    \begin{equation}
        L(\omega) = -\frac{1}{2}\mathbb{E}[(y_t - Q_\omega (s_t, a_t))^2],
    \end{equation}
   and its gradients with respect to the parameters $\omega$ by:
    \begin{equation}
        \nabla_\omega L (\omega) = -\mathbb{E}[(y_t-Q_\omega (s_t,a_t))\nabla_\omega Q_\omega(s_t, a_t)].
    \end{equation}
    We can then use a gradient-based optimizer to update the parameters in the direction that minimizes the loss. However, since the target values depend on the critic itself, this may quickly lead to instabilities in the training process. It is therefore common to use target models $Q'$ and $\pi'$ to calculate the target values $y_t$ \cite{Lillicrap2015_ddpg}. The parameters of these target models slightly lag behind the true models. For every iteration $i$ of training the critic, the target models are updated as:
    \begin{equation}
    \begin{split}
        \theta_{i}' = \tau \theta_{i} + (1-\tau)\theta_{i-1}' \\
        \omega_i' = \tau \omega_i + (1-\tau)\omega_{i-1}'.
    \end{split}
    \end{equation}
    Here, $\tau< 1$ is a hyper-parameter that determines how quickly the target models are updated.
    
    \subsection{Data collection}
    Updating the actor and critic requires evaluating expectation values over the state visitation distribution. This can naturally be done by collecting tuples ($s_t$, $a_t$, $r_t$ $s_{t+1}$) while following the control law $\pi_\theta$. For our purpose, this is done by running the AO system in closed-loop with the controller $\pi_\theta$. This data collection process is often separated into episodes of finite length, after which the sequence of tuples is saved in a so-called replay buffer, which is our training data set. For computational reasons, the expectation values are often estimated over a limited number of these tuples, a batch, instead of over the full data set. To reduce the correlation in a batch, the tuples are randomly sampled from the replay buffer.
    
    The DDPG algorithm is off-policy\cite{Silver_dpg,Lillicrap2015_ddpg}, meaning that it is not strictly required to sample the training data with the current policy. In principle, we can use any other controller, or even random actions, to generate the training tuples of states, actions and rewards. While this introduces a mismatch between the distribution of the training data and the true state visitation distribution of the current controller, this is often ignored in practice.
    Therefore, we can use previously collected data, where the control law was different, to train the controller. Furthermore, it is also possible to include data from a completely different controller, such as an integrator, to train the RL controller. This way, the algorithm could also make use of historically collected telemetry data using the current controller at the telescope.
    
    Because the algorithm allows for the data collection to be done with a different controller, we can also add some known disturbances to the commands. This is called exploration in the RL framework. For example, we can add some zero mean, normally distributed exploration noise with standard deviation $\sigma$:
    \begin{equation}
        a_t = \pi_\theta(s_t) + \mathcal{N}(0, \sigma) 
    \end{equation}
    This results in the controller not always giving the same DM commands for a given state, such that we can observe the results of these different commands. A common trick is to use large initial exploration noise and slowly decay it. This allows the algorithm to first observe the results of large changes in DM commands and finally observe the result of small changes in these commands. We empirically choose the standard deviation of the exploration noise during episode $k$ as:
        \begin{equation}\label{eq:exploration}
        \sigma_k = \frac{\sigma_0}{1+\zeta k} \; \textrm{rad}.
    \end{equation}
    Here, $\sigma_0$ is a hyper-parameter that specifies the initial standard deviation of the exploration noise and $\zeta$ how quickly it decays.

    \subsection{State representation} \label{sec:state_representation}

    It is crucial to have a representation of the state $s_t$ such that the problem obeys the Markov property (Eq. \ref{eq:markov}), which requires that the state-transition probabilities are fully described by the current state and the action taken. This is not trivial in closed-loop AO control. A first guess for the state might be the most recent wavefront observation, $o_t$. This may, for example, be retrieved from a dedicated wavefront sensor (WFS). However, defining the state like this is not consistent with the Markov property. First of all, we do not have access to the full state of the environment but only observe a noisy representation with the WFS measurements. Furthermore, the most recent wavefront measurement does not contain all the required information; for example, it lacks information about possible vibrations and wind flow. When the AO loop is closed, we also have to distinguish changes in the turbulence profile from changes due to previous commands, as we only observe the closed-loop residuals. These issues may be solved using state augmentation, where sequences of previous wavefront measurements and applied DM commands are used in the state. However, if we want our controller to be able to perform under varying conditions, it may require a large number of previous measurements to have a full representation of the state. For example, Guyon \& Males\cite{Guyon2017_eof} used a vector of the previous 800 measurements to correct tip-tilt vibrations. A disadvantage of this is that it increases the computational demand. Furthermore, all previous measurements are in principle weighted equally, which may lead to additional noise if the problem is not correctly regularized.
     
     To circumvent the problem of partial observability, it was shown that Recurrent Neural Networks (RNNs) are able to solve control problems even for partially observable Markov Decision Processes, where the Markov property is not obeyed \cite{Wierstra2007, Heess2015_rdpg}. RNNs are a type of neural network often used in processing sequential data\cite{Graves2013_RNN_speech, Goodfellow_deeplearning}. Instead of only mapping an input to an output, RNNs also update a memory state $m_t$ based on the previous memory state and the input. This allows RNNs to summarize previous information in this memory state and use it to calculate the output. This enables RNNs to learn an internal representation of the true state of the system based on sequences of noisy measurements. As an additional advantage calculating the control commands only requires the most recent observation to be propagated through the controller. This is computationally less demanding, which may be important for high-order AO systems operating at kHz frequencies. Furthermore, RNNs also effectively use the temporal structure of the data by incorporating basic priors, such as that the most recent measurement is the most relevant instead of using one large vector of components with equal weights. In our case, we will use a Long Short-Term Memory (LSTM) cell \cite{Hochreiter1997_lstm} as the recurrent architecture. Training is now done on sequences of states, actions and rewards using Truncated Backpropagation Trough Time (TBTT) over a length $l$. This version of the algorithm is known as the Recurrent Deterministic Policy Gradient (RDPG) algorithm \cite{Heess2015_rdpg}.
     
     The state for closed-loop AO now consists of only the most recent observation of the wavefront $o_t$ and the previous DM commands $a_{t-1}$:
     
     \begin{equation}
         s_t = (o_t, a_{t-1})\;.
     \end{equation}

    \subsection{Algorithm Overview}
\begin{figure}[htbp]
        \centering
        \includegraphics[width=0.9\linewidth]{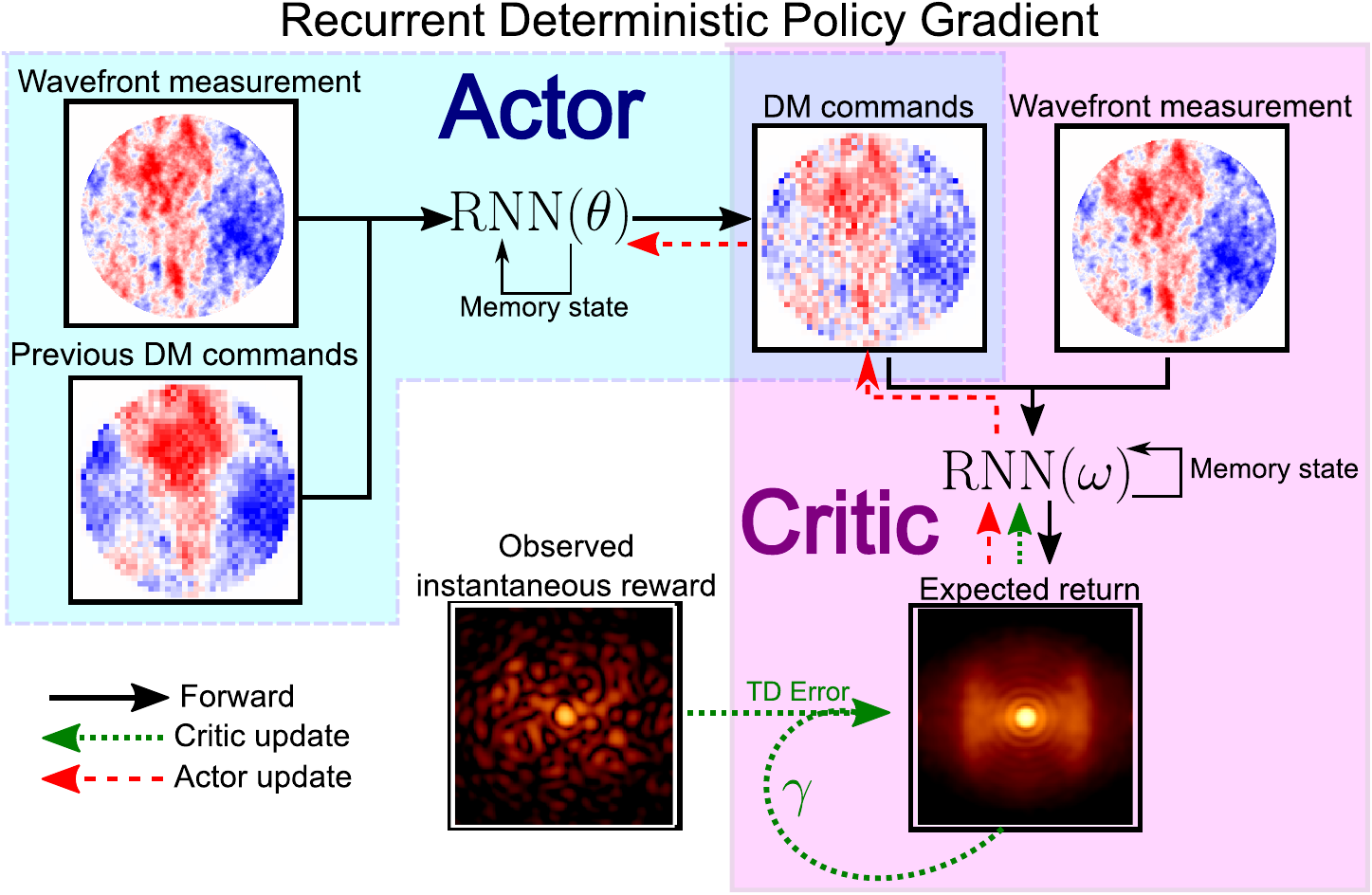}
        \caption{Visual overview of the algorithm. The black line indicates forward propagation. The green and red lines indicate how we backpropagate the gradients to update the critic and the actor, respectively.}
        \label{fig:rdpg}
    \end{figure}

    The algorithm can be summarized in three main steps that are iteratively performed:
    \begin{enumerate}
        \item Collect training sequences ($o_1,a_1,r_1,\dots o_{t}, a_t, r_t$) by running in closed loop and storing the data in the replay buffer.
        \item Train the critic using Temporal Difference learning on the observed data.
        \item Improve the control law using the gradients obtained from the critic.
    \end{enumerate}
    The pseudo-code describing the full algorithm can be found in Algorithm \ref{algo:rdpg}, and a schematic overview is shown in Fig \ref{fig:rdpg}.
    
    \RestyleAlgo{ruled}
    \begin{algorithm}[htbp]
     \caption{Recurrent Deterministic Policy Gradient for closed-loop AO (based on Heess et al. \cite{Heess2015_rdpg})}
      \label{algo:rdpg}
     Initialize critic parameters $\omega$ and actor parameters $\theta$ randomly\\
     Initialize target network parameters $\omega'=\omega$, $\theta'=\theta$ \\
     Initialize replay buffer \\
     Initialize $\sigma_0$ and $\zeta$ \\
     \For{k=1,number of episodes}{
       Reset controller memory states $m_t=0$ and deformable mirror \\
       \For{t=1,T}{
         Receive wavefront measurement $o_t$, reward $r_t$, construct state $s_t = (o_t, a_{t-1})$ and select incremental DM commands $a_t = \pi(s_t) + \epsilon$, with $\epsilon \sim \mathcal{N}(0,\sigma_k)$
      }
      Save trajectories $(o_1,a_1,r_1,\dots,o_T,a_T,r_T)$ in replay buffer \\
      Decay exploration noise $\sigma_k = \frac{\sigma_0}{1+\zeta k}$\\
      \For{j=1,number of training steps}{
      Randomly sample N batches of length $l$ from the replay buffer \\
      For each batch $i$ construct state histories $h_t^i = ( (o_1^i, a_0^i) \dots, (o_t^i, a_{t-1}^i))$ \\
      Calculate the critic targets using the Bellman equation: $$y_t^i = r_t^i + \gamma Q'_{\omega'}(h_{t+1}^i,\pi'_{\theta'}(h_{t+1}^i)) $$\\
      Calculate the sampled critic loss gradient with TBTT:
      $$ \nabla_\omega L(\omega) \approx \frac{1}{Nl} \sum_{i=1}^N \sum_{t=1}^l (y_t^i - Q_\omega(h_t^i,a_t^i))\nabla_\omega Q_\omega(s_t^i,a_t^i) $$\\
      Calculate the sampled Policy Gradient with TBTT:

      $$\nabla_\theta J(\theta) \approx \frac{1}{Nl} \sum_i^N \sum_t^l \nabla_{a} Q_\omega(h_t^i,a_t^i) \nabla_\theta \pi_\theta(h_t^i) $$\\
      Use a gradient-based optimizer (e.g. Adam) to update the actor parameters $\theta$ and critic parameters $\omega$.\\
      Update the target network parameters:
      \begin{align*}
        \theta_j' &= \tau \theta_j + (1-\tau)\theta_{j-1}' \\
        \omega_j' &= \tau \omega_j + (1-\tau) \omega_{j-1}' \\
      \end{align*}

      }
      }
    \end{algorithm}
    
\section{Tip-tilt control in simulations} \label{sec:tip_tilt_simulations}
    \subsection{Simulation setup}
    As a proof of concept we apply the algorithm to tip-tilt control in simulations. We simulate an idealized optical system using the HCIPy \cite{por2018hcipy} package for python. Our system has an unobscured, circular aperture of 1 meter in diameter and operates at a wavelength of 1~$\mu$m. The simulated deformable mirror has only two degrees of freedom, its tip and tilt. The observation $o_t$ is the center of gravity ($x_t, y_t$) of the focal plane image in units of $\lambda/D$. The measured center of gravity along with the previous DM commands $a_{t-1}$ are fed into a controller that controls tip and tilt in closed loop. To represent the servo-lag of estimating the wavefront tip and tilt, calculating and applying the control commands, we delay the DM commands by a discrete number of frames:
    \begin{equation}
        \textrm{DM}_t = \textrm{DM}_{t-1} + a_{t-\tau},
    \end{equation}
    where $\tau$ is the servo lag. This is equivalent to the controller seeing the state from $\tau$ iterations ago, which implies a delay in obtaining the reward. Throughout this section we assume an AO system operating at 1 kHz with a servo lag of 3 frames and do not consider any detector noise. Furthermore, we only simulate tip-tilt errors and ignore all higher order modes in the simulations.
    
    \subsubsection{Algorithm setup} \label{sec:hyperparams}
    To minimize the residual tip-tilt jitter we choose the squared deviation of the center of gravity of the PSF as the reward function:
    \begin{equation}
        r_t = -\frac{x_t^2+y_t^2}{b^2}.
    \end{equation}
    Here, $b$ is a scaling factor for which we use a value of $4 \lambda/D$ in the simulations. This scaling factor acts as a normalization of the rewards. Here, we empirically choose a value for the scaling factor, but this could also be inferred from the data. Important to note is that this scaling factor also scales the gradient of the reward with respect to the DM commands and therefore also influences the gradient update of the actor. 
    
    We use the same Neural Network architecture for both the actor and the critic. The input consists of the wavefront measurement $o_t= (x_t, y_t)$ and the previously applied tip-tilt commands $a_t = (a_{x,t-1}, a_{y,t-1})$. We use a Long Short-Term Memory \cite{Hochreiter1997_lstm} (LSTM) cell with 64 neurons as the recurrent layer in both the actor and the critic. For the critic the DM commands $a_t$ are appended to the output of the LSTM. After that, we have a fully connected layer with 64 neurons for both networks. Finally, the output of the actor consists of two neurons, the additional tip and tilt commands, while the output of the critic is a single neuron that estimates the expected future return $Q(s_t, a_t)$. The LSTM uses a tanh activation function for the input and output gates and a hard sigmoid for the recurrent gates while the fully connected layer uses the ReLU activation function. The output of the actor again uses a tanh activation function to constrain the incremental DM commands between -1 rad and +1 rad. The output of the critic has a linear activation function. The architectures are also listed in Table \ref{tab:arch_tiptilt}. Both the actor and critic have $\sim22,000$ free parameters. This is likely a lot more than required for tip-tilt control. Optimization of the architecture may result in shorter training times and improved performance, but is beyond the scope of this work.
    
    We use the gradient-based Adam optimizer algorithm \cite{Kingma2015_adam} with default parameters except for the learning rate. After every 500 iterations (an episode), we reset the DM shape. To reduce the correlation of the data in a batch early on, we only start training after a certain number of episodes, the warmup. We use a warmup of 5 episodes, which is equivalent to 2500 iterations, and the exploration law given in Eq. \ref{eq:exploration}. The hyperparameters of the RDPG algorithm are given in Table \ref{tab:hyperparams}. The algorithm and Neural Networks are implemented using a combination of the TensorFlow\cite{tensorflow2015-whitepaper} version 1.15 and Keras\cite{chollet2015keras} packages for Python.
    
\newcommand{\STAB}[1]{\begin{tabular}{@{}c@{}}#1\end{tabular}}
\begin{table}[htbp]
\center
\caption{Neural Network architectures of the actor and critic for the tip-tilt control.}
\label{tab:arch_tiptilt}
\begin{tabular}{l|l|c|c|c}
                                       \hline      & Layer type  & Neurons & Input shape                                                                 & Activation function \\ \hline
\multicolumn{1}{c|}{\multirow{3}{*}{\STAB{\rotatebox[origin=c]{90}{\large Actor}}}}  & LSTM             & 64      & (4)                                                              & Tanh                \\
\multicolumn{1}{c|}{}                        & Dense                & 64       & (64)                                                             & ReLU                \\
\multicolumn{1}{c|}{}                        & Dense                & 2       & (64)                                                              & Tanh             \\ \hline \\ \hline
 & Layer type   & Neurons & Input shape                                                                 & Activation function \\ \hline
\multicolumn{1}{l|}{\multirow{4}{*}{\STAB{\rotatebox[origin=c]{90}{\large Critic}}}} & LSTM             & 64      & (4)                                                              & Tanh                \\
\multicolumn{1}{l|}{}                        & Concatenate & -                & (64) and  (2) & -                   \\
\multicolumn{1}{l|}{}                        & Dense                 & 64      & (66)                                                             & ReLU              \\
\multicolumn{1}{l|}{}                        & Dense                & 1       & (64)                                                              & Linear             
\end{tabular}
\end{table}

    \begin{table}[htbp]
  \center
\caption{Hyperparameters}
\label{tab:hyperparams}
\begin{tabular}{c|c}
\hline
Parameter                         & Value          \\ \hline
Actor learning rate               & $10^{-5}$      \\
Critic learning rate              & $10^{-3}$      \\
Target network soft update $\tau$ & $10^{-3}$      \\
Discount factor $\gamma$          & $0.99$         \\
Batch size                        & 64             \\
TBTT length $l$                   & 50 ms            \\
Initial exploration $\sigma_0$      & 0.3 rad       \\
Exploration decay $\zeta$         & 0.005         \\
Episode length                    & 500 iterations \\
Number of training steps per episode& 500 \\
\end{tabular}
\end{table}
    
    \subsection{Vibration Suppression}
\begin{figure}[htbp]
            \centering
            \includegraphics[width=\linewidth]{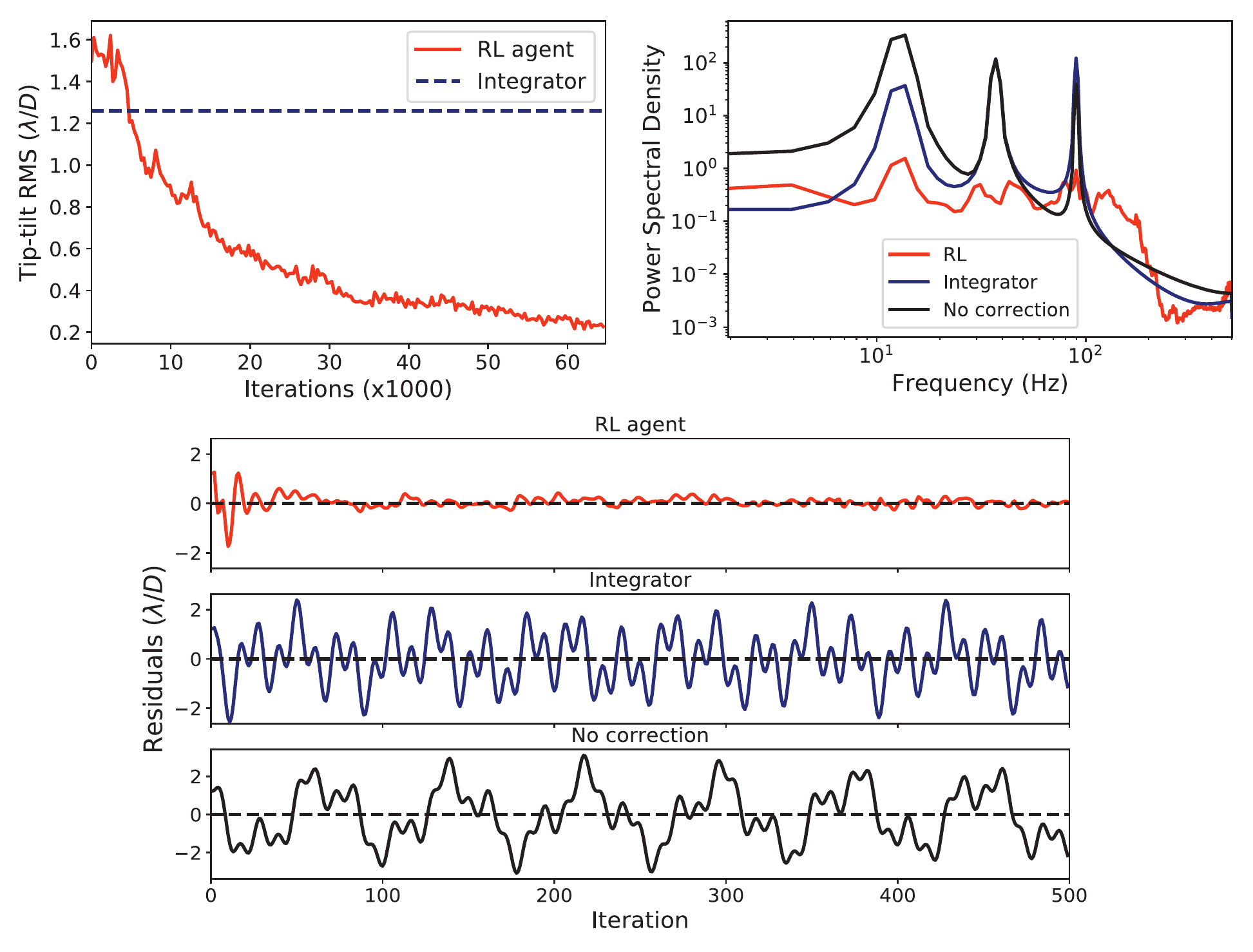}
            \caption{\textbf{Top left:} Training curve of the Reinforcement Learning controller under input disturbance consisting of three vibrations along both directions. Also shown is the average performance of an integrator with optimal gain. \textbf{Top right:} Temporal PSD of the residuals in the $x$-direction for a simulation of 10 seconds. Note that the lines of the input spectrum and integrator overlap at the peak at 37 Hz. \textbf{Bottom:} Residuals for the different controllers in the $x$-direction for a 500 ms simulation.}
            \label{fig:vibrations}
        \end{figure}

    As an example we consider an input disturbance consisting of three pure vibrations along each of the $x$ and $y$ directions. Along the $x$ direction we have vibrations with frequencies at 13 Hz, 37 Hz and 91 Hz and along the $y$ direction at 11 Hz, 43 Hz and 87 Hz. The gain of the integrator is optimized by running in closed loop for 1 second and choosing the gain that gives the lowest residual root mean square (RMS) center deviation:
    \begin{equation}
        \textrm{RMS} = \sqrt{<x_t^2+y_t^2>}.
    \end{equation}
    Figure \ref{fig:vibrations} shows the evolution of the RMS during the training of the RL controller. It shows that after a few thousand iterations the RL controller outperforms the integrator and eventually reaches an average RMS that is a factor ~6 lower than for the integrator. The figure also shows the temporal PSD of the residuals along the $x$-direction for the fully trained controller without exploration noise. This PSD is estimated using Welch's method\cite{Welch1975} for a simulation of 10 seconds. The integrator reduces the power of the vibration at 13 Hz, but the other vibrations fall outside its correction bandwidth, even adding power to the 91 Hz vibration. The RL controller almost completely removes the power from all three vibrations. Most remaining power is at a frequency of 13 Hz. This may be explained by the fact that the period of this vibration is longer than the optimization length of 50 ms. The residuals in the $x$-direction for a 500-ms simulation for the RL controller, integrator and without correction are shown in Fig. \ref{fig:vibrations}. We see that after a short time of initializing the memory states, the RL controller is able to mitigate the vibrations.
        {}
    
    \subsection{Power-law Disturbances}
\begin{figure}[htbp]
            \centering
            \includegraphics[width=\linewidth]{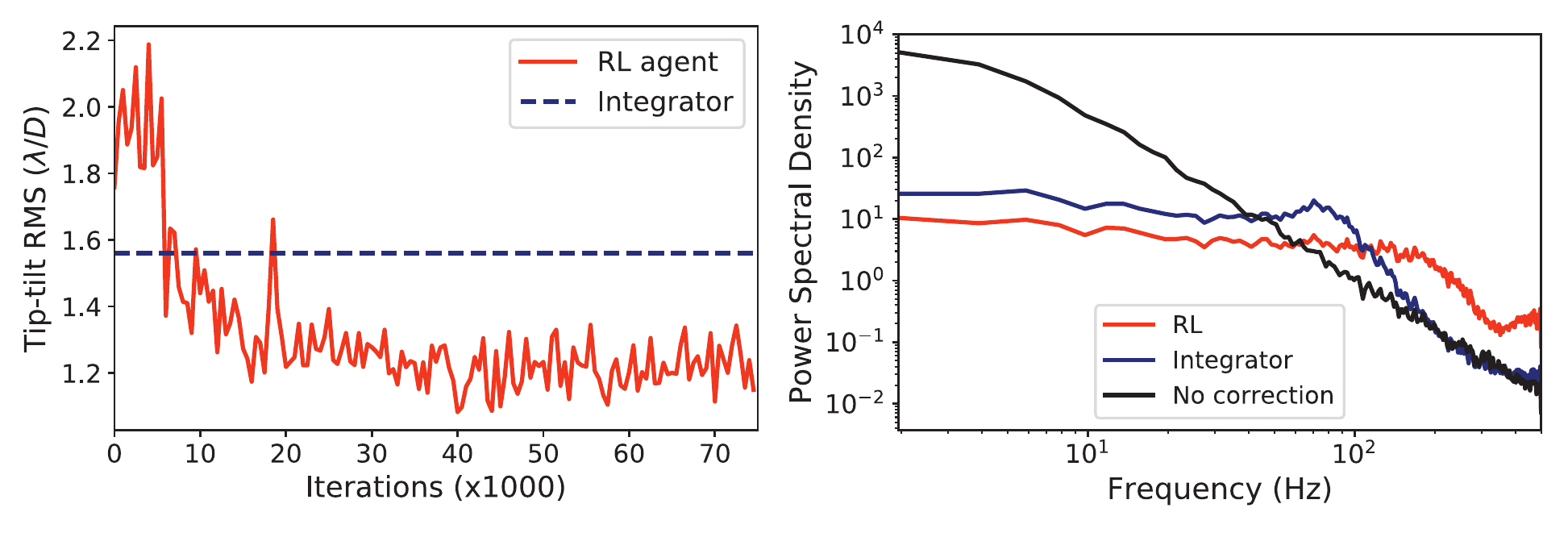}
            \caption{\textbf{Left:} Training curve of the Reinforcement Learning controller under power-law input turbulence and the average performance of an integrator with optimal gain. \textbf{Right:} Temporal PSD of the residuals in the x-direction for a simulation of 10 seconds.}
            \label{fig:power_law}
        \end{figure}

    Next, we consider input disturbances following a temporal power law with a power-law index of -8/3 along both directions. We chose this power-law index because phase fluctuations for Kolmogorov turbulence are described by a -8/3 power-law\cite{Conan1995_temporal}. For every episode, we generate random power-law time series for 500 ms, neglecting frequencies below 2 Hz. The training curve is shown in Fig. \ref{fig:power_law}. We again observe better RMS performance for the RL controller as compared to an integrator with optimized gain. We test the trained controller by running in closed loop for 10 s and calculating the temporal PSD of the residuals, which are also shown in Fig. \ref{fig:power_law}. The RL controller has a larger rejection power at lower frequencies. The residual PSD of the integrator exhibits a bump around ~70 Hz, which is the result of the integrator overshooting. A lower gain would reduce this overshooting but would also decrease its rejection power at lower frequencies. The RL controller can account for already applied commands that are not yet visible in the residual measurements due to the servo lag, allowing it to use a higher gain without overshooting. The RL controller adds power at high frequencies. This is the result of the waterbed effect, which implies that if we decrease the power at one part of the spectrum this always leads to the amplification of another part of the spectrum.
    
        {}

\section{Lab verification of tip-tilt control}\label{sec:tip_tilt_lab}
 \subsection{Experimental Setup}
\begin{figure}[htbp]
        \centering
        \includegraphics[width=0.9\linewidth]{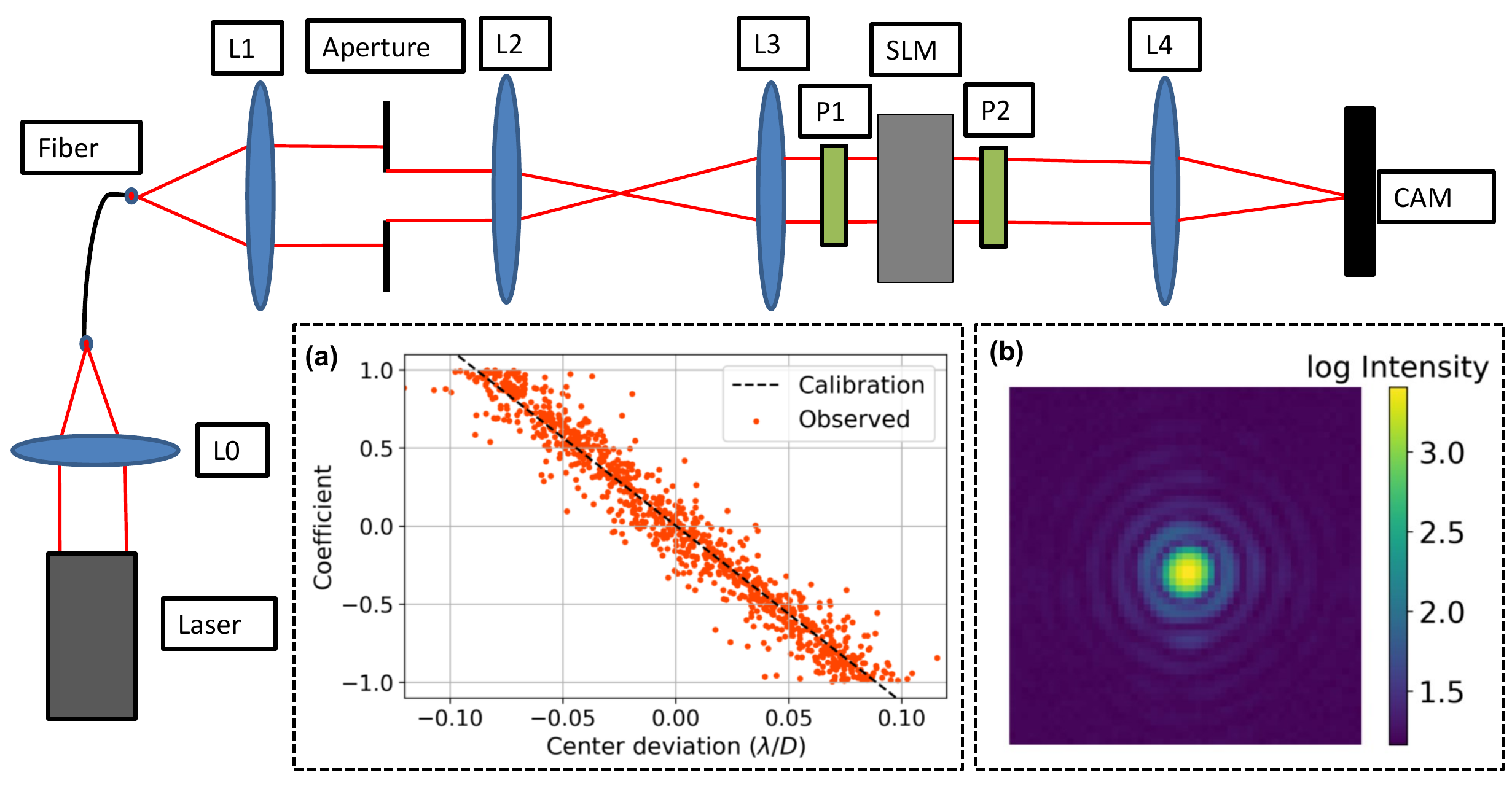}
        \caption{Lab setup to test the RL control algorithm with lenses L0-L4, Polarizers P1 and P2, a Spatial Light Modulator (SLM) and a camera (CAM). Insets: \textbf{(a)} Measured tilt in the focal plane as a function of the applied phase gradient on the SLM. \textbf{(b)} Measured Point Spread Function for the setup.}
        \label{fig:lab_setup}
    \end{figure}

The results in the previous section are based on idealized simulations. To validate our results for tip-tilt control we tested the algorithm in the lab. The lab setup is sketched in Figure \ref{fig:lab_setup}. The first lens focuses a 633-nm He-Ne laser onto a single-mode fiber. The fiber output is then collimated, and a diaphragm defines the input aperture. A 4f system reimages the aperture onto a Holoeye LC2002 Spatial Light Modulator (SLM) with 800x600 pixels. The pupil covers approximately 200x200 pixels on the SLM. Right before the SLM we define the polarization with a polarizer. After a second polarizer we focus the beam onto the camera. The resulting Point Spread Function is shown in Figure \ref{fig:lab_setup}.

We use the SLM as both the turbulence generator and the corrector. The polarizers are rotated such that the SLM mainly manipulates the phase profile with minimum amplitude modulation. The voltages of the SLM are controlled with 8-bit resolution. As the response for both small and large voltages becomes highly nonlinear, only values between 25 and 225 are used. Tip and tilt aberrations are introduced by applying a gradient on the SLM at the location of the pupil. Figure \ref{fig:lab_setup} shows the calibration of the SLM for the resulting shift of the PSF in the focal plane image. The coefficient refers to the amplitude of the gradient profile, where 1 corresponds to the maximum possible gradient in the pupil. The maximum shift is approximately $\pm$0.1 $\lambda/D$. The refresh rate of the SLM is 30 Hz according to the manufacturer. We use the SLM at a frequency of approximately 18 Hz. We have attempted to measure the delay of the control loop but this appeared to not be constant. This might be the result of varying processing speeds in the pipeline. On average, there is a delay of 0.1 to 0.15 seconds, which is equivalent to 2-3 frames at 18 Hz. For the RL algorithm we use the same architectures, hyperparameters and reward function as in the simulations (see Section \ref{sec:hyperparams}), except for the reward scaling $b$, for which we now use a value of 0.05 $\lambda/D$, because the shift in the PSF is smaller than in the simulations.

\subsection{Vibration Suppression}
\begin{figure}[htbp]
        \centering
        \includegraphics[width=\linewidth]{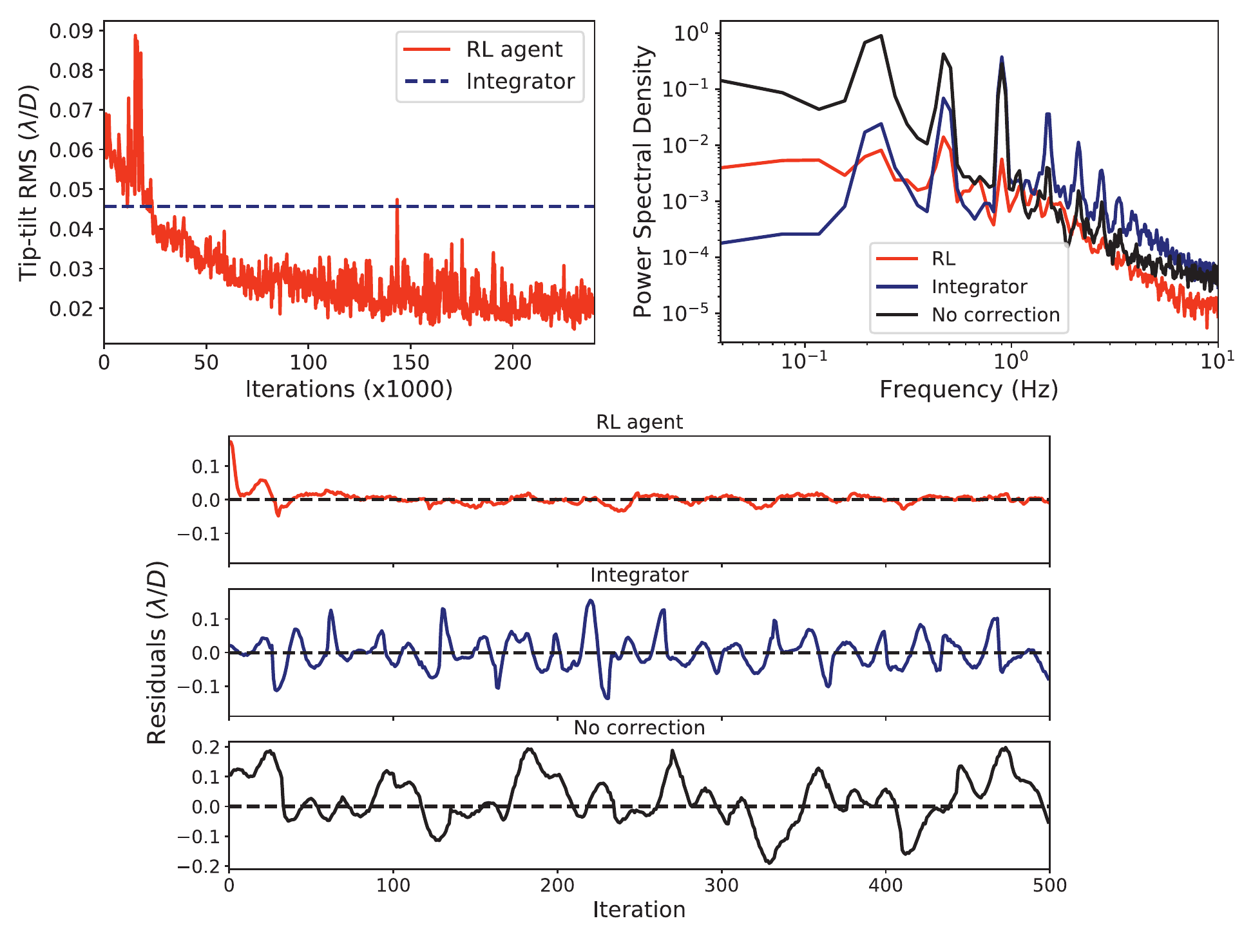}
        \caption{\textbf{Top left:} Training curve of the Reinforcement Learning controller for an input spectrum consisting of 3 vibrations along each direction in the lab. For comparison, we show the average performance of an integrator with optimal gain. \textbf{Top right:} Temporal PSD of the residuals in the x-direction for a measurement of 2500 iterations. \textbf{Bottom:} Residuals for the two controllers in the $x$-direction for a 500-iteration measurement. }
        \label{fig:lab_vibrations}
    \end{figure}

We again test the ability of the algorithm to filter out a combination of 3 vibrations along each direction. We insert vibrations at 0.23, 0.48 and 0.9 Hz in the $x$-direction and at 0.18, 0.51 and 0.86 Hz in the $y$-direction. Figure \ref{fig:lab_vibrations} shows the training curve and the residual temporal PSD for the trained agent and an integrator with optimized gain. It also shows the residuals for a 500-iteration measurement. These results confirm our findings in the simulations as we again see a much reduced power in the vibrations. There is a little more power left in the vibrations as compared to the simulations, and the training time is longer. This is likely the result of the additional noise in the lab, variable control-loop delay and dynamics of the SLM. 
    
\subsection{Identification and mitigation of a varying vibration}\label{sec:varying_vib}
\begin{figure}[htbp]
    \centering
    \includegraphics[width=\linewidth]{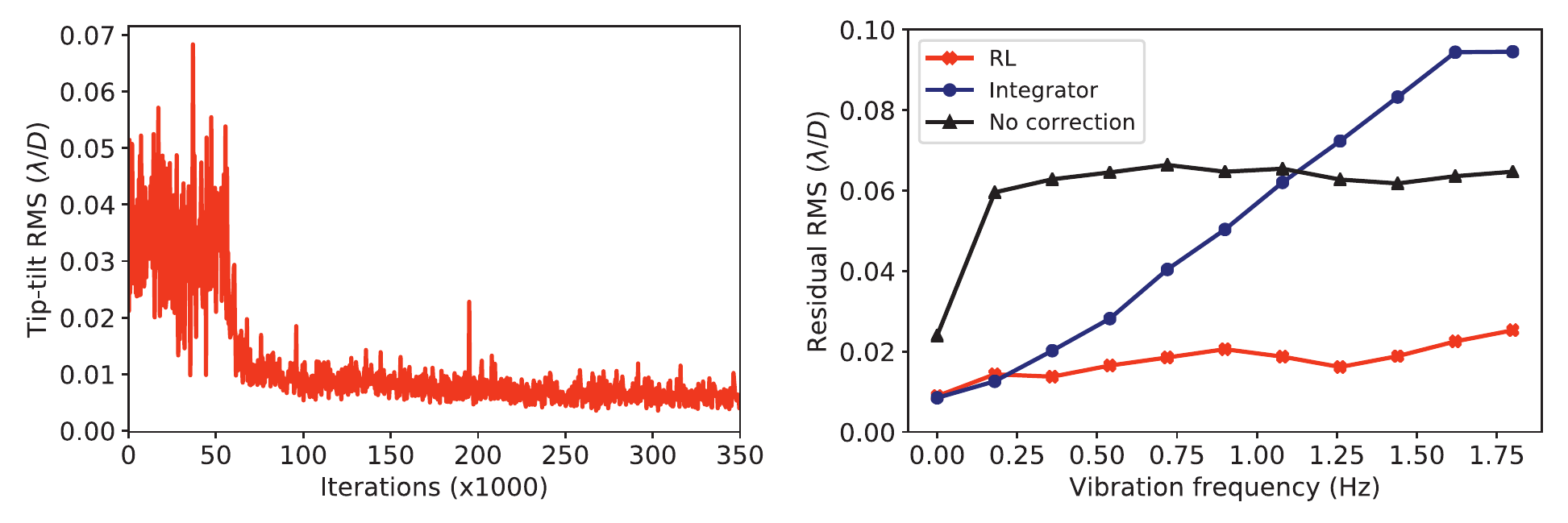}
    \caption{\textbf{Left:} Training curve of the Reinforcement Learning controller for a single vibration with random amplitude between 0 and 1 and random frequency between 0 and 1.8 Hz. \textbf{Right:} Residual RMS as a function of vibration frequency for a 2000-iteration measurement.}
    \label{fig:varying_vib}
\end{figure}

So far, we have only considered disturbances with a stationary power spectrum. Once the input spectrum changes (e.g. the frequency of the vibrations), we would have to retrain the controller. However, Recurrent Neural Networks should be able to identify the relevant parameters online, without retraining. To test this, we train the controller with a single but randomly varying input vibration along the $x$-axis. After every 500 iterations we reset the vibration parameters: the amplitude is randomly sampled between 0 and 1 and the frequency between 0 and 1.8 Hz. The training curve is shown in Fig. \ref{fig:varying_vib}. Once it has converged there is very little variance in the RMS, meaning that the performance is mostly independent of the amplitude and frequency of the input vibration. We compare the performance of the trained controller to that of an integrator with a fixed gain of 0.6. We apply a vibration with a fixed amplitude of 1 and a specific frequency and run closed loop for 2000 iterations. The residual RMS as a function of the frequency of the vibration is shown in Fig. \ref{fig:varying_vib}. The data point for a frequency of 0 Hz is without an input vibration and is only due to natural turbulence on our optical table. The RL controller is able to reduce the RMS for all frequencies. On the other hand, the integrator amplifies the vibrations starting at a frequency of 1.25 Hz. This is the result of the vibration falling outside the correction bandwidth for this gain. However, the RNN is able to adaptively change its commands based on the observed sequence of measurements. This demonstrates that the RNN can learn to identify the relevant parameters of the disturbance, in this case the amplitude and frequency of the vibration, without updating the control law.

\subsection{Power-law Disturbances}
\begin{figure}[htbp]
        \centering
        \includegraphics[width=\linewidth]{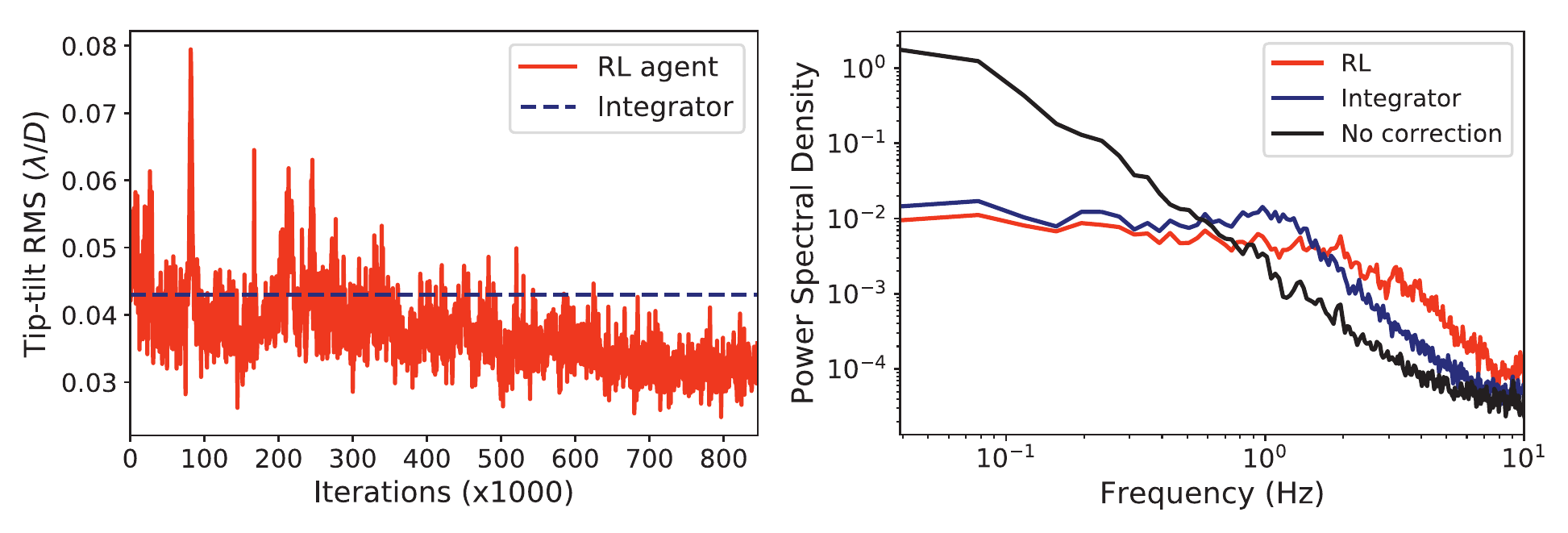}
        \caption{\textbf{Left:} Training curve of the Reinforcement Learning controller under power-law input turbulence in the lab along with the average performance of an integrator with optimal gain. \textbf{Right:} Temporal PSD of the residuals in the x-direction for a measurement of 2500 iterations.}
        \label{fig:lab_power}
    \end{figure}

We also tested the performance for a power-law input spectrum. The training curve and residual temporal PSD are shown in Fig. \ref{fig:lab_power}. It takes many iterations before it converges. This is likely because of the noisy response of the SLM. We expect that with the improvements suggested in Section \ref{sec:full_wavefront}, this training time could be significantly reduced. Comparing the residual PSD to that of an integrator with optimized gain, we see results that are similar to the simulations, although the improvement is smaller. We again observe an increased rejection power at low frequencies without the bump due to overshooting.

      {}

\section{Full wavefront control}
\label{sec:full_wavefront}

\subsection{Simulation setup}
We simulate a typical high-contrast imaging instrument with an unobscured aperture with a diameter of 8 meters and a DM with 41x41 actuators operating at a loop frequency of 1 kHz. All simulations are again done using HCIPy\cite{por2018hcipy}. We use a noiseless wavefront sensor, which directly senses the phase and project this onto the DM. Therefore, the control basis consists of Gaussian influence functions centered at the location of the DM actuators. This actuator basis has the advantage that the spatial response of each mode is the same but at a different location in the pupil. This results in each mode having the same spatio-temporal wavefront correlations, except for the actuators located at the edges. It also retains the spatial structure of the wavefront. The effective loop delay between sensing the wavefront and the application of the DM commands is 2 ms. To estimate the effect on the contrast, we propagate the residual wavefront through an ideal coronagraph \cite{Guyon2006_coronagraph}. The parameters of the simulations are shown in Table \ref{tab:sim_params}.

\begin{table}[htbp]
\caption{Parameters of the AO simulations.}
\label{tab:sim_params}
\center
\begin{tabular}{l|l}
\hline
Parameter              & Value     \\ \hline
Aperture diameter      & 8 m       \\
Wavelength             & 1 $\mu m$ \\
Loop frequency         & 1 kHz     \\
Servo-lag              & 2 ms      \\
Number of illuminated actuators & 1201 
\end{tabular}
\end{table}

\subsection{Architecture \& reward}
\begin{figure}[htbp]
    \centering
    \includegraphics[width=0.35\linewidth]{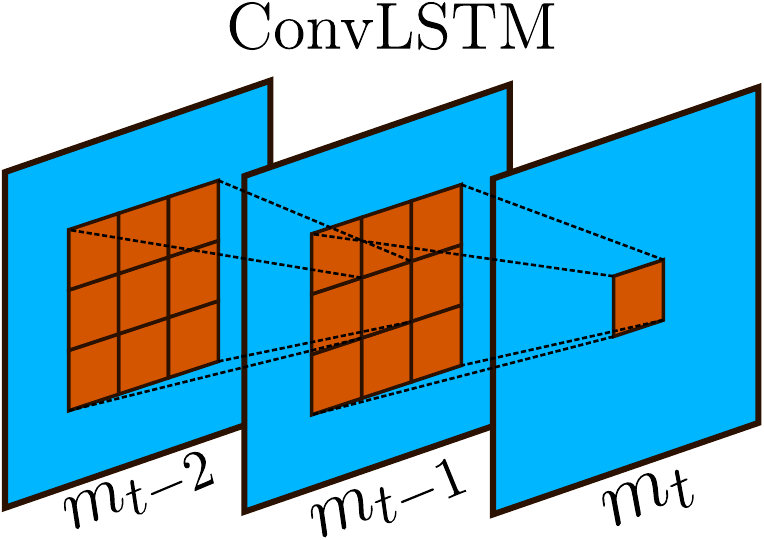}
    \caption{Illustration of the recurrent part of a Convolutional LSTM layer. The output at time $t$ is determined by a local field of view of the memory states $m_{t-1}$ and the local inputs.}
    \label{fig:convlstml}
\end{figure}

To take advantage of the spatio-temporal structure of the data we combine Recurrent Neural Networks and Convolutional Neural Networks (CNNs). CNNs are commonly used in image classification and regression problems \cite{Krizhevsky2012_alexnet, he2015_resnet} and assume that features are local and translationally invariant to drastically reduce the number of free parameters as compared to regular ANNs. In these CNNs, kernels are convolved with the input data to obtain the activation in the next layer. In our case this means that we assume that each actuator has the same spatial response, and only close-by actuators are important for the prediction in the next step. We use a Convolutional Long Short-Term Memory (ConvLSTM) layer \cite{Shi2015_convlstm}. This is equivalent to a LSTM for each actuator with a local field of view and shared weights between the actuators. The recurrent part of the Convolutional LSTM is illustrated in Figure \ref{fig:convlstml}. In the case of a 1x1 kernel size, we would have the same decoupled control law for each individual actuator. However, this would not allow the algorithm to capture any spatial correlations in the data. Although these spatial correlations are found to be small in on-sky telemetry data for wind velocities $<$10 m/s on short timescales\cite{VanKooten2020_PC}, they may become more important for higher wind speeds or longer timescales. Furthermore, they may help in identifying the direction of the wind flow or to mitigate noise. 

As the reward, we would ideally want to directly optimize the Strehl ratio or the post-coronagraphic contrast in the science image. However, the accurate estimation of these quantities at kHz frequencies could be difficult. Therefore we use the squared residual wavefront error as the reward. This residual wavefront error is calculated from the WFS measurements. A disadvantage of this is that the reward suffers from the same limitations as the WFS. For example, it means that the algorithm is insensitive to modes to which the WFS is blind and suffers from the same aliasing effects as the WFS. Instead of using a single scalar as reward, we use a 2D reward map, with the squared residual error of each actuator mode at the actuator's location. This allows us to use a fully convolutional model for the critic, where only close-by actuators are considered when estimating the loss. This results in less free parameters and the access to more and localized information, leading to decreased training times as opposed to a global metric.

The input to our controller is the residual wavefront errors $o_t$ and previous incremental DM commands $a_t$ at the location of the actuator, concatenated along the depth, the number of inputs for each actuator. This means the input is an array of shape 41x41x2 for each time step. For unused actuators outside the pupil, we fill the array with zeros. Our actor architecture consists of a Convolutional LSTM layer with 16 3x3 kernels. After that, we have two regular convolutional layers with 8 and 1 3x3 kernels, respectively, which provide the new incremental DM commands $a_t$. For the critic, we again have a Convolutional LSTM layer with 16 3x3 kernels. After that we concatenate the DM commands to the output of the Convolutional LSTM. This is followed by three regular convolutional layers with 16, 8 and 1 3x3 kernels, respectively. The output of the critic is a 41x41x1 map of the expected return for each actuator. All convolutional layers use the tanh activation function except for the last layer of the Critic, which uses a linear activation function. The architectures are summarized in Table \ref{tab:arch_full}. Both models have $\sim$12,000 free parameters. Again, we have not done an optimization of the architecture or hyperparameters.

\subsubsection{Algorithm improvements}
Instead of starting with a randomly initialized control law, as in our tip-tilt control, we pre-train the actor using supervised learning to mimic an integral controller with a sub-optimal gain of 0.4. This slightly reduces the training time. In addition, we add regularization to the DM commands to penalize the controller for potentially adding unsensed modes. We do this by adding an $L_2$ penalty term on the norm of the incremental DM commands to the loss function:
\begin{equation}
    J(\theta) = \mathbb{E}[R]-\frac{1}{2}\lambda \mathbb{E}[a^2],
\end{equation}
where $\lambda$ is the strength of the regularization. The new policy gradient then becomes:
\begin{equation}
    \nabla_\theta J(\theta) = \mathbb{E}[(\nabla_a Q_\omega(s,a)-\lambda a) \nabla_\theta\pi_\theta(s)].
\end{equation}

Finally, we slightly alter the training procedure. Before doing the truncated backpropagation through time, we first propagate the 10 previous states through the models. This ensures that the memory states are initialized correctly. Initialization was less of an issue for the tip-tilt control as the TBTT length was longer and therefore the initial timesteps contributed less to the total gradient as opposed to full AO control.

The hyperparameters for the algorithm used throughout this section are listed in Table \ref{tab:hyperparams_full}. A notable change compared to the tip-tilt control is the lower discount factor $\gamma$ because the optimization horizon does not need to be long. This is because of the monotonic nature of the reward in the absolute wavefront error; a command that decreases the wavefront error in the short term is also the best in the long-term. We still have to use $\gamma>0$ because of the delay in the commands and reward. We also use a much higher target soft update $\tau$ and noise decay $\zeta$, which both lead to decreased training times without giving up too much stability.

\begin{table}[]
\caption{Neural Network architectures of the actor and critic used for the full wavefront control.}
\label{tab:arch_full}
\center
\begin{tabular}{l|c|c|c|c|c}
                                \hline             & Layer type  & Kernel size & Kernels & Input shape                                                                 & Activation function \\ \hline
\multicolumn{1}{c|}{\multirow{3}{*}{\STAB{\rotatebox[origin=c]{90}{\large Actor}}}}  & ConvLSTM    & 3x3         & 16      & (41, 41, 2)                                                              & Tanh                \\
\multicolumn{1}{c|}{}                        & Conv        & 3x3         & 8       & (41, 41, 16)                                                             & Tanh                \\
\multicolumn{1}{c|}{}                        & Conv        & 3x3         & 1       & (41, 41, 8)                                                              & Tanh              \\ \hline \\ \hline
 & Layer type  & Kernel size & Kernels & Input shape                                                                 & Activation function \\ \hline
\multicolumn{1}{l|}{\multirow{5}{*}{\STAB{\rotatebox[origin=c]{90}{\large Critic}}}} & ConvLSTM    & 3x3         & 16      & (41, 41, 2)                                                              & Tanh                \\
\multicolumn{1}{l|}{}                        & Concatenate & -           & -       & (41, 41,16),  (41,41, 1) & -                   \\
\multicolumn{1}{l|}{}                        & Conv        & 3x3         & 16      & (41, 41, 17)                                                             & Tanh                \\
\multicolumn{1}{l|}{}                        & Conv        & 3x3         & 8       & (41, 41, 16)                                                             & Tanh                \\
\multicolumn{1}{l|}{}                        & Conv        & 3x3         & 1       & (41, 41, 8)                                                              & Linear             
\end{tabular}
\end{table}

    \begin{table}[htbp]
  \center
\caption{Hyperparameters used for the full wavefront control }
\label{tab:hyperparams_full}
\begin{tabular}{c|c}
\hline
Parameter                         & Value          \\ \hline
Actor learning rate               & $10^{-5}$      \\
Critic learning rate              & $10^{-3}$      \\
Target network soft update $\tau$ & $0.1 $      \\
Discount factor $\gamma$          & $0.9$         \\
Batch size                        & 32             \\
TBTT length $l$                   & 20 ms             \\
Action regularization $\lambda$    & $10^{-3}$            \\
Initial exploration $\sigma_0$      & 0.5 rad       \\
Exploration decay $\zeta$         & 0.1         \\
Episode length                    & 1000 iterations \\
Number of training steps per episode& 1000 \\
\end{tabular}
\end{table}

\subsection{Stationary turbulence}\label{sec:full_stationary}
\begin{figure}[htbp]
    \centering
    \includegraphics[width=\linewidth]{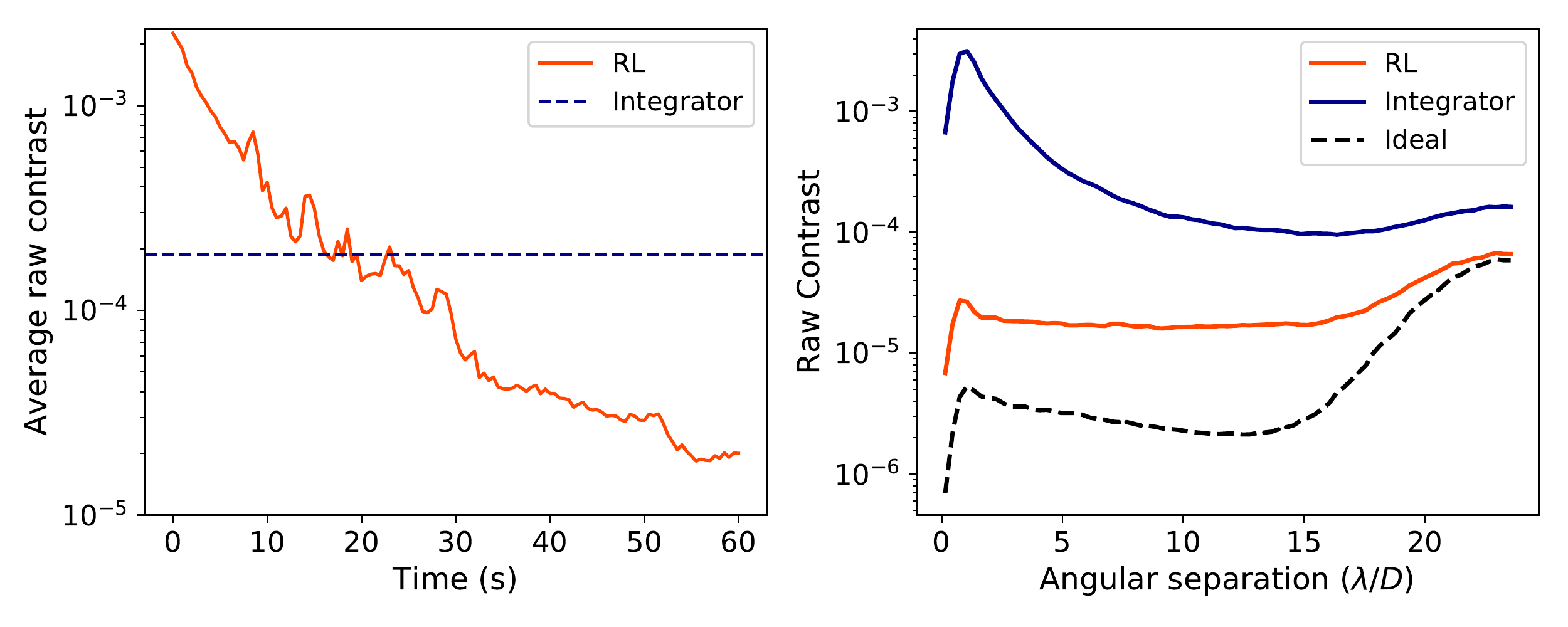}
    \caption{\textbf{Left:} Evolution of the average raw contrast after an ideal coronagraph within the control radius during the training of the RL controller for turbulence consisting of two layers of frozen-flow turbulence. \textbf{Right:} Raw contrast curves for a 10-second integration of the focal plane for the simulations with a two-layer atmosphere. }
    \label{fig:contrast_stationary}
\end{figure}

\begin{figure}[htbp]
    \centering
    
    \includegraphics[width=\linewidth]{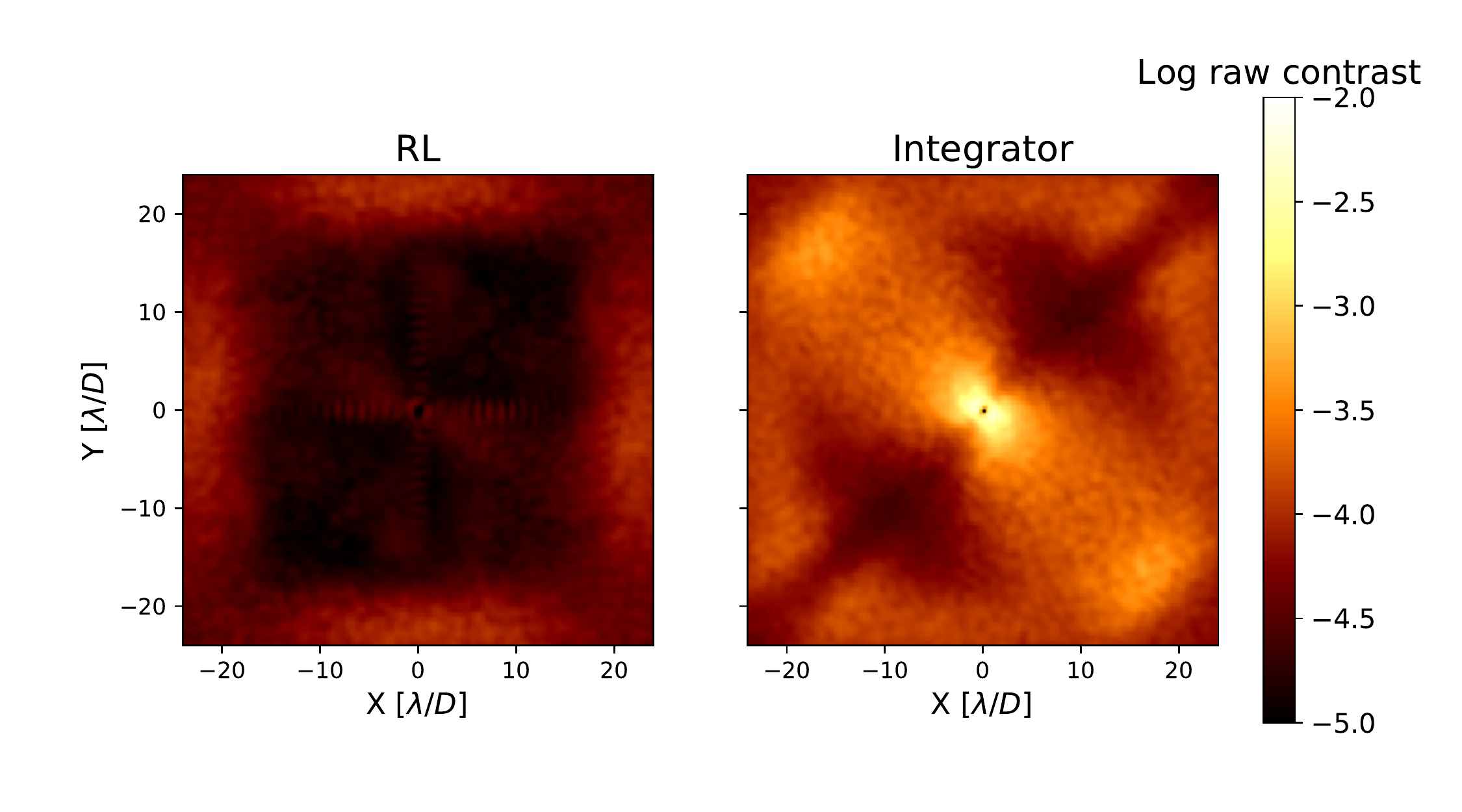}
    \caption{10-second integration of the focal plane after an ideal coronagraph for the simulations with an atmosphere consisting of two layers of frozen-flow turbulence. }
    \label{fig:image_stationary}
\end{figure}

First, we test the algorithm for input turbulence with stationary parameters. We simulate an atmosphere consisting of two layers of frozen-flow turbulence: A ground layer with a horizontal wind speed of 10 m/s with an $r_0$ of 15 cm at 500 nm and a jet stream layer with a wind speed of 30 m/s at an angle of 45 degrees with an $r_0$ of 20 cm at 500 nm. Here, an angle of 0 degrees corresponds to the turbulence flowing from left to right across the pupil. The average raw contrast behind the ideal coronagraph within the control radius during the training is shown in Figure \ref{fig:contrast_stationary}. After training we test the algorithm without exploration noise. Figure \ref{fig:image_stationary} shows the resulting coronagraphic focal plane image for a 10-second simulation for our RL controller and an integrator with optimized gain. The gain of the integrator was optimized by running in closed loop for 3 seconds to determine the gain corresponding to the largest Strehl ratio. We find an optimal gain of 0.55 for the integrator. Figure \ref{fig:contrast_stationary} shows the radially averaged raw contrast profile for these images. The raw contrast is calculated as the average intensity in that radial bin divided by the Strehl ratio of the non-coronagraphic focal plane image. We observe an increase in raw contrast of up to two orders of magnitudes at small separations as compared to the integrator. Also shown is the contrast curve for an ideal controller, which perfectly removes all wavefront errors that can be fitted with the DM at each timestep, without any delay.

\subsection{Non-stationary turbulence}
\begin{figure}[htbp]
    \centering
    \includegraphics[width=0.55\linewidth]{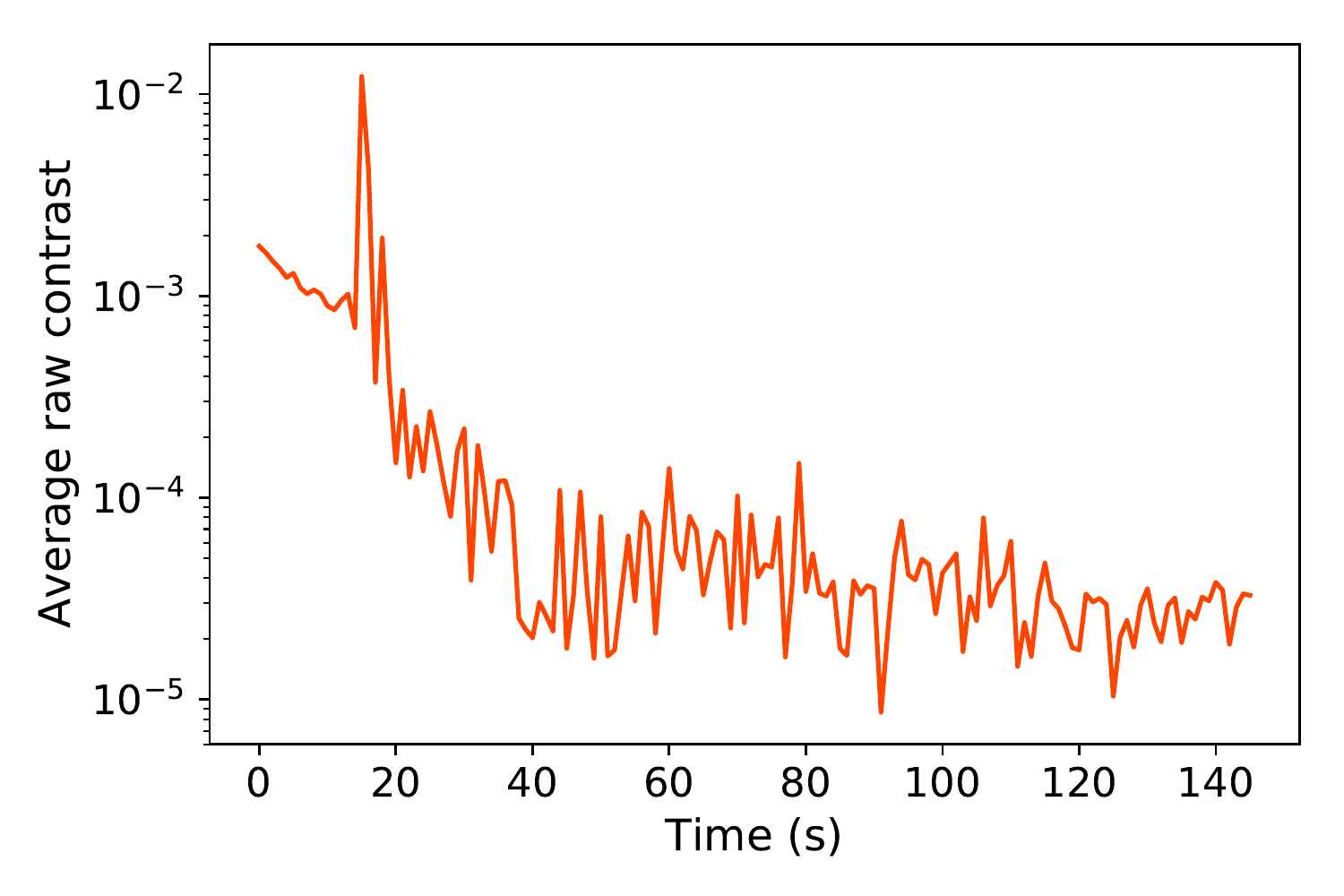}
    \caption{Evolution of the average raw contrast during the training phase behind an ideal coronagraph within the control radius for an atmosphere containing of a single layer of frozen-flow turbulence with random wind speed and direction.}
    \label{fig:training_curve_random_wind}
\end{figure}

\begin{figure}[htbp]
    \centering
    \includegraphics[width=\linewidth]{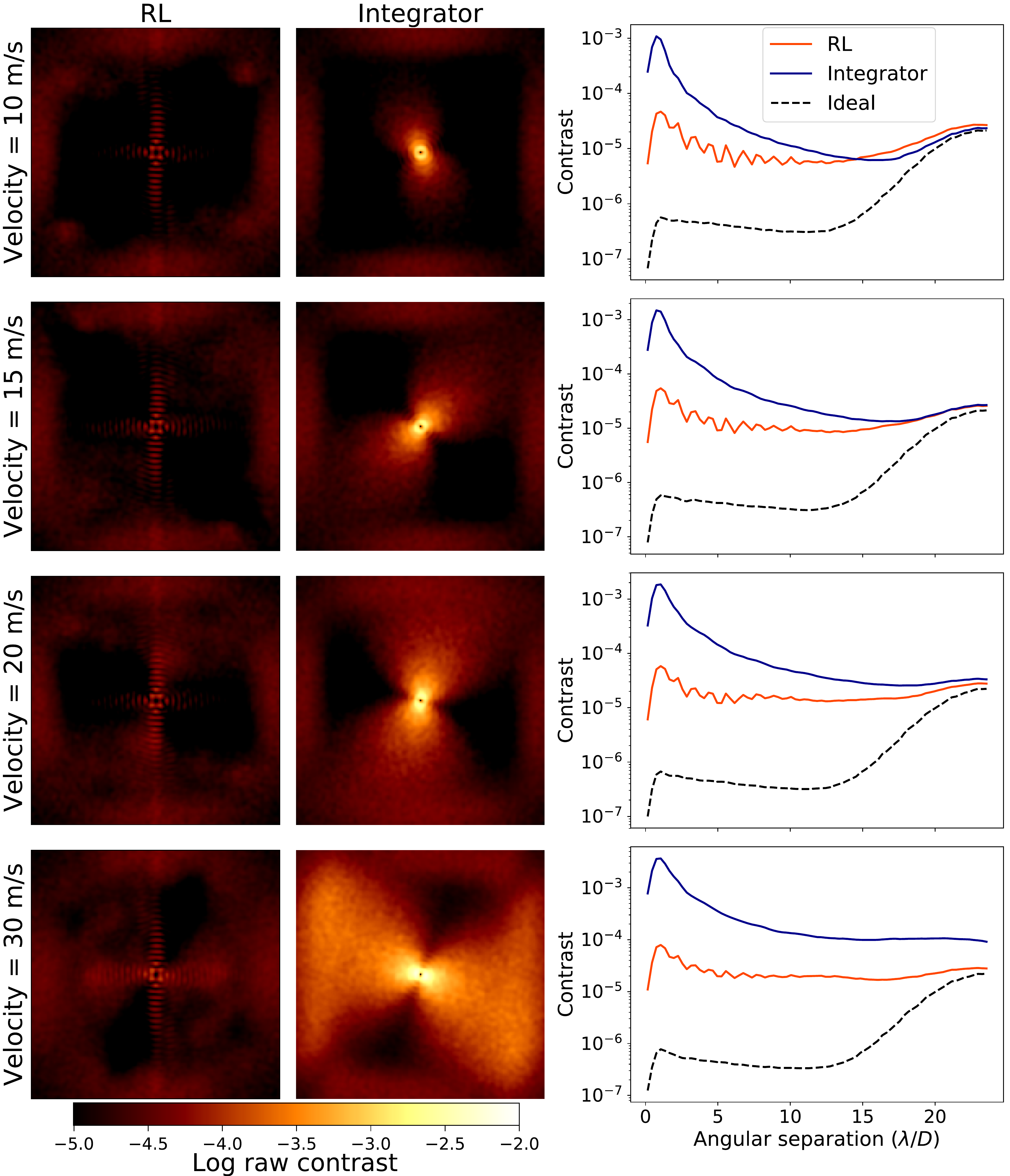}
    \caption{Resulting coronagraphic focal-plane images and contrast curves for a 10 second simulation for a specific wind velocity and a random wind direction. Each row represents a given wind velocity, which is listed on the left. The left column shows the image using the Reinforcement Learning controller, the middle column the image obtained while using an integrator with optimized gain and the right column shows the resulting contrast curves.}
    \label{fig:images_varying_wind}
\end{figure}

One may wonder if the algorithm can learn to adapt to changing turbulence conditions on its own. To explore this, we train the algorithm under a variety of conditions. We randomly reset the wind speed and direction every second. The wind speed is uniformly sampled between 10 and 40 m/s, and we allow all angles for the wind direction. Only a single layer of frozen-flow turbulence is used in these simulations. The training curve is shown in Figure \ref{fig:training_curve_random_wind}. After training, we test the algorithm for specific wind speeds and randomly sampled directions. The resulting focal plane images and contrast curves for a 10-second simulation are shown in Figure \ref{fig:images_varying_wind}.

\begin{figure}[htbp]
    \centering
    \includegraphics[width=0.7\linewidth]{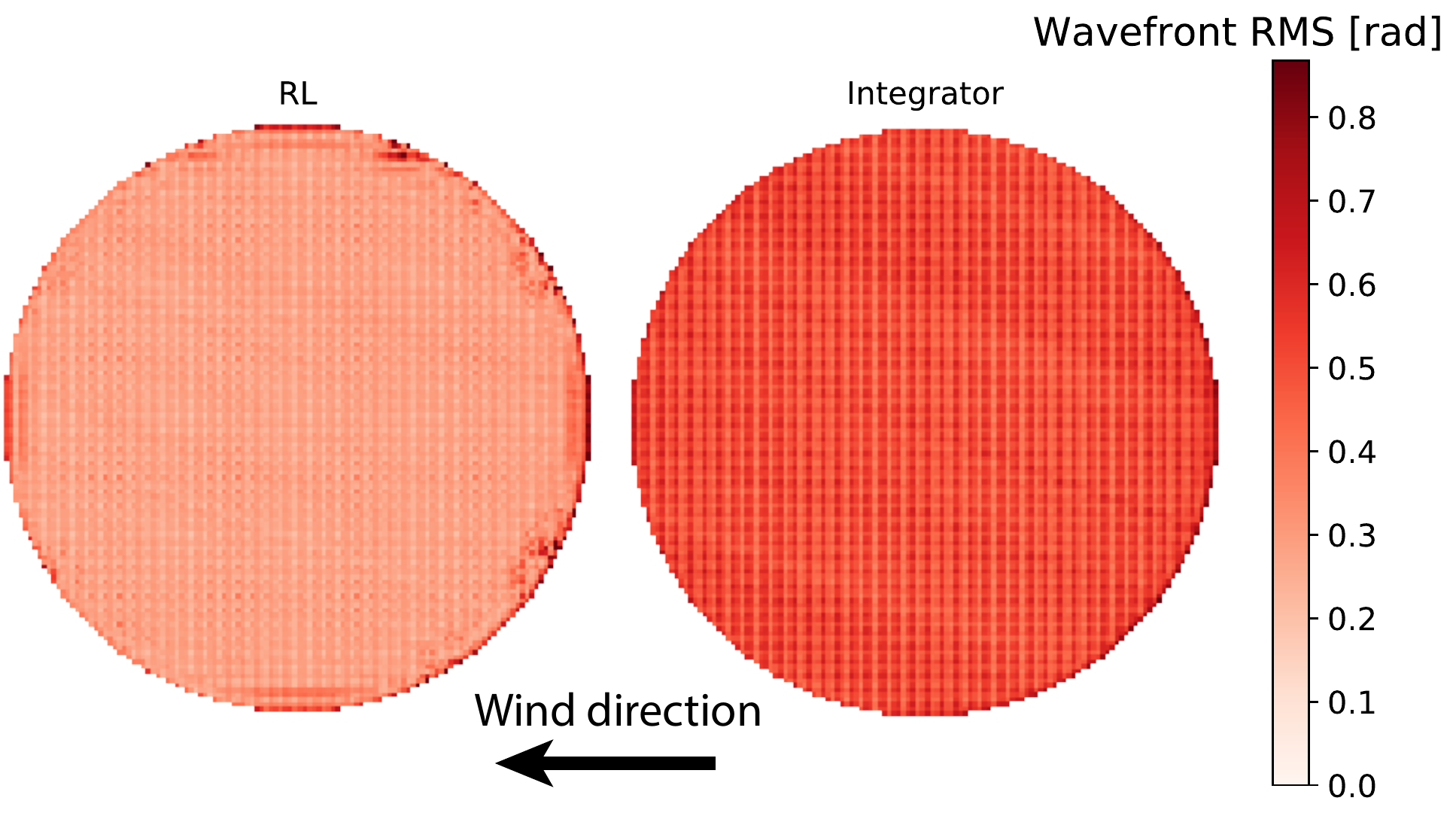}
    \caption{Residual wavefront RMS for a wind speed of 30 m/s with the controller trained on random wind directions.}
    \label{fig:wf_rms}
\end{figure}

Our results indicate that the controller improves the contrast for all wind velocities and directions in these simulations. This demonstrates that the controller is able to learn to adapt to changes in the spatio-temporal PSD as a result of a different wind speed or angle. The controller therefore does not necessarily need online learning, in contrast to most other predictive control algorithms. However, the improvement in contrast is less than when optimized for a fixed wind speed and angle, as some wind-driven halo remains. This is because of the finite representation power of the model; a deeper or wider model will likely perform better, at the expense of added computational cost. Furthermore, this generalization would be even harder in the case of more complex turbulence with more varying parameters. An alternative approach may therefore be to start off with the general model and learn online for the current observing conditions. This may significantly reduce the time needed to train the controller. In the same way as adapting to different wind speeds, we also expect the algorithm to be able to adapt to different $r_0$ and account for the nonlinearities accordingly when using a Pyramid Wavefront Sensor. There is a residual cross-shaped ripple pattern in the coronagraphic images when using the RL controller. This is the result of square-edge effects, as the assumption that all actuators have the same spatial response does not hold for actuators at the edge. This can be seen in Fig. \ref{fig:wf_rms}, which shows the residual wavefront RMS for a simulation with a wind speed of 30 m/s. One solution to this might be to not have all convolutional kernels have shared weights, at the cost of more free parameters.

\section{Conclusions \& Outlook}\label{sec:conclusions}
    In conclusion, Reinforcement Learning is a promising approach towards data-driven predictive control for adaptive optics. We have shown how a closed-loop Recurrent Neural Network (RNN) controller can be trained using the Recurrent Deterministic Policy Gradient algorithm without any prior knowledge of the system dynamics and disturbances. This RNN only needs the most recent wavefront measurement and the previous DM commands at every time step. 
    
    First, we applied the algorithm to tip-tilt control for a simulated, ideal AO system. We demonstrated that our approach can learn to mitigate a combination of tip-tilt vibrations, reducing the residual RMS by a factor of $\sim6$ as compared to an optimal-gain integrator. We also show a $\sim 25\%$ decrease in residual RMS for power-law input turbulence compared to the optimal gain integrator. These experiments were then repeated in a lab setup. For the vibration mitigation we observe a decrease in RMS by a factor 2.2 and a $\sim 25\%$ decrease in residual RMS for power-law input turbulence. Furthermore, we showed that a single Recurrent Neural Network controller can mitigate a vibration with varying amplitude and frequency without needing online updating of the control parameters.
    
    Secondly, we showed in simulations how the algorithm can be applied to the control of a high-order DM. We showed that for an atmosphere consisting of a ground layer and a jet-stream layer, the algorithm can improve the contrast at small separations by two orders of magnitude as compared to an optimal-gain integrator. Furthermore, when the controller is trained over a variety of observing conditions, it is able to adapt to changes in the wind speed and direction without needing online updating of the control law. This relaxes the real-time computational demand that would be required to update the control law given the constantly changing atmospheric conditions. It also simplifies the implementation because control and learning can be decoupled. However, the obtained contrast is lower than when trained for stationary conditions. A deeper or wider model may improve its ability to generalize and perform under varying conditions, at the expense of added computational time.
    
    We have not considered the effect of photon noise on the performance of the algorithm. As the noise affects both the state and the reward, we expect that this will increase the amount of data needed to train the controller. Future research should test the effect of noise on the performance of the controller. The algorithm should also be tested with a nonlinear wavefront sensor, such as the pyramid wavefront sensor, because our approach has the advantage of easily accommodating a nonlinear control law. Furthermore, the algorithm is off-policy, meaning that we can train on data that is not collected with the current optimal control law. We can therefore use historical telemetry data collected using an integrator to train the controller. It should be investigated if this can already provide a good controller or if we need to learn online.
    
    Although we have shown the potential of Reinforcement Learning for closed-loop predictive control, there are a few more practical considerations. First is the complexity of the implementation of the algorithm on real-time controllers (RTC) at actual telecopes, because current RTCs are often not compatible with the Deep Learning interfaces used in this paper. Another consideration is the computational cost. Although prediction is easily within the capabilities of current RTCs because of the partially decoupled approach, online learning with backpropagation will be computationally very challenging. For example, the full training process, including optical simulations and gradient updates, for the full wavefront control in the stationary case (see Section \ref{sec:full_stationary}) took about 4 hours using the NVIDIA Tesla K80 GPU offered by Google Colab. While this can be sped up significantly by the use of multiple GPU's or specialized hardware, training times are likely to increase under more complex turbulence and with real noise. However, we have shown that the algorithm can learn to perform under varying conditions, allowing control and training to be decoupled, making the computational cost of training less of an issue. Alternatively, one could use the generalized model and finetune it for the current observing conditions, significantly reducing the training time. Another consideration is the stability of the learning algorithm and controller. Since we are using a sophisticated nonlinear controller, it is difficult to give performance guarantees. Furthermore, changing the hyperparameters can significantly influence the learning stability of the algorithm. 
    
\subsection*{Acknowledgements}
We wish to thank the reviewers for their feedback, which has resulted in improvements of this work. R.L. acknowledges funding from the European Research Council (ERC) under the European Union's Horizon 2020 research and innovation program under grant agreement No 694513. Support for this work was provided by NASA through the NASA Hubble Fellowship grant \#HST-HF2-51436.001-A awarded by the Space Telescope Science Institute, which is operated by the Association of Universities for Research in Astronomy, Incorporated, under NASA contract NAS5-26555. Part of this work was already presented in Ref.  \citenum{Landman2020_RL}.

\subsection*{Disclosures}
The authors declare no conflict of interest.

\bibliography{report}   
\bibliographystyle{spiejour}   


\vspace{2ex}\noindent\textbf{Rico Landman} is a PhD candidate at Leiden Observatory working on direct imaging and spectroscopy of exoplanets. He received his BS degrees in physics and astronomy and MS degree in astronomy from Leiden University in 2018 and 2020, respectively.

\vspace{2ex}\noindent\textbf{Sebastiaan Y. Haffert} is a NASA Hubble Postdoctoral Fellow at the University of Arizona's Steward Observatory. His research focuses on high-spatial and high-spectral resolution instrumentation for exoplanet characterization.

\vspace{2ex}\noindent\textbf{Vikram M. Radhakrishnan} is a PhD candidate at Leiden Observatory working on innovative control approaches to high-contrast imaging.

\vspace{2ex}\noindent\textbf{Christoph U. Keller} is a Professor of Experimental Astrophysics at Leiden Observatory. He specializes in developing innovative optical instruments for astronomy, remote sensing and biomedical imaging.


\listoffigures
\listoftables

\end{document}